\documentclass[aps,twocolumn,pra,superscriptaddress,showpacs]{revtex4}
\usepackage{amssymb}
\usepackage{mathrsfs}
\usepackage{graphicx}
\usepackage{amsmath}
\usepackage{multirow}
\usepackage{hhline}

\begin{document}

\begin{titlepage}

\title
{A qubit strongly-coupled to a resonant cavity: asymmetry of the spontaneous emission
spectrum beyond the rotating wave approximation}
\author{Xiufeng Cao\footnote{Email: xfcao@xmu.edu.cn}}
\affiliation{Advanced Science Institute, RIKEN, Wako-shi 351-0198, Japan}
\affiliation{Department of Physics and Institute of Theoretical
Physics and Astrophysics, Xiamen University, Xiamen, 361005,China}
\author{J. Q. You}
\affiliation{Advanced Science Institute, RIKEN, Wako-shi 351-0198,
Japan} \affiliation{Department of Physics and state Key Laboratory of Surface Physics,
Fudan University, Shanghai
200433, China}
\author{H. Zheng}
\affiliation{Department of Physics, Shanghai Jiao Tong University,
Shanghai 200240, China}
\author{Franco Nori}
\affiliation{Advanced Science Institute, RIKEN, Wako-shi 351-0198,
Japan} \affiliation{Physics Department, The University of Michigan,
Ann Arbor, Michigan 48109-1040, USA}
\date{\today }

\begin{abstract}
We investigate the spontaneous emission spectrum of a qubit in a
lossy resonant cavity. We use neither the rotating-wave
approximation nor the Markov approximation. The qubit-cavity
coupling strength is varied from weak, to strong, even to lower bound of the ultra-strong.
For the weak-coupling case, the spontaneous emission spectrum of the
qubit is a single peak, with its location depending on the spectral
density of the qubit environment. Increasing the qubit-cavity
coupling increases the asymmetry (the positions about the qubit
energy spacing and heights of the two peaks) of the two spontaneous
emission peaks (which are related to the vacuum Rabi splitting)
more. Explicitly, for a qubit in a low-frequency intrinsic bath, the
height asymmetry of the splitting peaks becomes larger, when the
qubit-cavity coupling strength is increased. However, for a qubit in
an Ohmic bath, the height asymmetry of the spectral peaks is
inverted from the same case of the low-frequency bath, when the
qubit is strongly coupled to the cavity. Increasing the qubit-cavity
coupling to the lower bound of the ultra-strong regime, the height asymmetry of the left
and right peak heights are inverted, which is consistent with the
same case of low-frequency bath, only relatively weak. Therefore,
our results explicitly show how the height asymmetry in the
spontaneous emission spectrum peaks depends not only on the
qubit-cavity coupling, but also on the type of intrinsic noise
experienced by the qubit.
\\


\end{abstract}


\maketitle

\end{titlepage}

\section{Introduction}

Strong and ultra-strong qubit-cavity interactions have been achieved in both
cavity QED and circuit QED systems (see, e.g., \cite%
{phys-today,nature-458-178,nature-2-81,add1}). This opens up many new
possible applications. For example, one could use the cavity as a quantum
bus to couple widely-separated qubits in a quantum computer \cite{sa,gr}, as
a quantum memory to store quantum information, or as a generator and
detector of single microwave photons for quantum communications \cite{add1}.

As a demonstration of strong interaction in cavity QED and circuit QED
systems, the vacuum Rabi splitting has been an exciting subfield of optics
and solid-state physics (see, e.g., \cite{add2,add4,pra-69-062320,add5}),
after its observation in atomic systems \cite{prl-68-1132}. In 2004, two
groups \cite{nature-432-197,nature-432-200}\ reported the experimental
realization of vacuum Rabi splitting \ in semiconductor systems: a single
quantum dot in a spacer of photonic crystal nanocavity and in semiconductor
microcavity, respectively. In the same year, the experiment \cite%
{nature-431-162} showed that the vacuum Rabi splitting can also been
obtained in a superconducting two-level system, playing the role of an
artificial atom, coupled to an on-chip cavity consisting of a
superconducting transmission line resonator. When the qubit was resonantly
coupled to the cavity mode, it was observed \cite{nature-431-162} that two
well-resolved spectral lines were separated by a vacuum Rabi frequency $\nu
_{{\mathrm{Rabi}}}\approx 2g.$ Except for the asymmetry in the height of the
two split energy-peaks (e.g., Ref.~[\onlinecite{nature-431-162}]), the data
is in agreement with the transmission spectrum numerically calculated using
the rotating wave approximation (RWA). When considering the vacuum Rabi
splitting behavior for strong qubit-cavity coupling, the anti-rotating terms
should be taken into account, and this might explain the observed \textit{%
asymmetric} spontaneous emission (SE) spectrum.

\subsection{Antirotating terms are important for strong coupling QED}

It is obvious that the anti-rotating terms are not important when the
coupling between a qubit and the cavity field is sufficiently weak, and when
the energy spacing of the qubit is resonant with the central frequency of
the cavity. However, in the \textit{ultra-strong coupling} regime, the
anti-rotating terms of the intrinsic bath of the qubit coupling play an
important role \cite{prl-98-103602-2007,pra-80-033846,add8,arkiv}. As a
consequence of the anti-rotating terms in the Hamiltonian of cavity QED and
circuit QED, even the ground state of the system contains a finite number of
virtual photons. Theoretical research \cite{prl-103-147003,pra-80-053810}
reveal that these virtual photons can be released by a non-adiabatic
manipulation, where the Rabi frequency $g(t)$ is modulated in time at
frequencies comparable or higher than the qubit transition frequency. This
phenomenon, called \textquotedblleft emission of the quantum vacuum
radiation\textquotedblright , would be completely absent if these
anti-rotating terms are neglected. The energy shift of the qubit in its
intrinsic bath has been studied in \cite{pre-80-021128-2009} using the full
description, (i.e., non-Markov and without RWA) and found that the
deviations from the previous approximation result already amount to $\sim
5\% $ for $g/\Delta \sim 0.1.$

In doped semiconductor quantum wells embedded in a microcavity (e.g., \cite%
{pra-74-033811}), considering the anti-rotating coupling of the intracavity
photonic mode and the electronic polarization mode, but using RWA in the
coupling to their respective environments, it was found \cite{pra-74-033811}
that for a coherent photonic input, signatures of the ultra-strong coupling
have been identified in the asymmetric and peculiar anticrossing of the
polaritonic eigenmodes. From the descriptions given above, it can be seen
that, as $g/\Delta $ increases to the ultra-strong coupling case, the
anti-rotating terms that are otherwise negligible become more relevant and
will lead to a profound modification in the nature of the quantum state of
the qubit system.

\subsection{The asymmetry of the two splitting Rabi peaks can be explained
beyond the RWA approximation}

In this paper, we study the SE spectrum of a qubit in a cavity. Our
calculations include \textit{two kinds of anti-rotating terms}: one from the
intrinsic qubit environment and the other one from the cavity environment.
This method is a powerful tool to investigate various kinds of
qubit-environment interaction with anti-rotating terms and without using the
Markov approximation. Because in this method, the qubit-environment coupling
terms higher than the two-order are droped, it constricts that this method
is unavailable in the case when the qubit-environment coupling strength is
larger than the energy spacing of the qubit \cite{add24g,add25s}. Comparing
the cases of a qubit in an Ohmic bath with the case of a qubit in a
low-frequency bath, we find that for the case of a qubit in a low-frequency
bath, as the qubit-cavity coupling strength increases, the height asymmetry
of two splitting peaks is enhanced. However, for the case of a qubit in an
Ohmic bath, the height asymmetry of the spectral peaks are inverted from the
same case of the low-frequency bath, when the qubit is strongly coupled to
the cavity. Increasing the qubit-cavity coupling to ultra-strong regime, the
height asymmetry of the left and right peaks are inverted, which is
consistent with the same case of low-frequency bath, only relatively weak.
Since experiments reported that \textit{a superconducting qubit intrinsic
bath is mainly due to low-frequency noise, our results are consistent with
experimental data }using a superconducting qubit in Ref.~[%
\onlinecite{nature-431-162}].

We also investigate the dependence of the SE spectrum on the strength of the
qubit-cavity coupling and the quality factor $Q$ of the cavity in either an
Ohmic or in a low-frequency intrinsic qubit bath. Furthermore, we
distinguish the contributions to the asymmetry from each bath: the intrinsic
qubit bath and the cavity bath, and clarify the reason for the different
kinds of peaks asymmetry. All of these results directly indicate that in the
strong coupling regime, the SE spectrum is deeply influenced by the
anti-rotating terms and the type of intrinsic noise experienced by the qubit.

\section{Beyond the Rotating Wave Approximation}

By using a cavity to confine the electromagnetic field, the strength of the
qubit-cavity interaction can be increased by several orders of magnitude to
the regime of strong or even ultra-strong coupling \cite{add7}. The
strong-coupling regime for cavity quantum electrodynamics has been reached
for natural atoms in optical cavities, superconducting qubits in circuit
resonators (i.e., on-chip cavities), and quantum dots in photonic-crystal
nanocavities. Recently, the ultra-strong coupling regime has be achieved for
a superconducting qubit in an on-chip cavity \cite{nature-458-178}. Although
the coupling of the qubit to the cavity is much stronger than the coupling
of the qubit to its intrinsic environment, the parameters in Ref.~[%
\onlinecite {nature-431-162}] show that both the decay rate of the cavity
photon ($\kappa /2\pi \approx 0.8$ $\mathrm{MHz}$) and the qubit decoherence
rate ($\gamma /2\pi \approx 0.7$ $\mathrm{MHz}$) are comparable. Therefore,
we model the environment of the qubit in a cavity using two bosonic baths:
one, called the intrinsic bath of the qubit and represented by operators $%
b_{k}$ and $b_{k}^{\dagger }$, is related to the relaxation of the qubit
induced by its intrinsic environment; and the other, denoted as cavity bath
of qubit and represented by the operators $a_{k}^{\dagger }$ and $a_{k},$
involves the relaxation of the qubit caused by photons in the cavity. Figure
1 schematically shows the model considered here. For the intrinsic qubit
bath, a broad frequency spectrum (e.g., either an Ohmic or a low-frequency
spectrum) can be used to characterize it. For the cavity bath, because of
the cavity leakage, it can be described by a Lorentzian spectrum with a
central frequency, i.e., a single-mode cavity with its frequency broadened
by the cavity leakage.
\begin{figure}[b]
\includegraphics[width=8cm,clip]{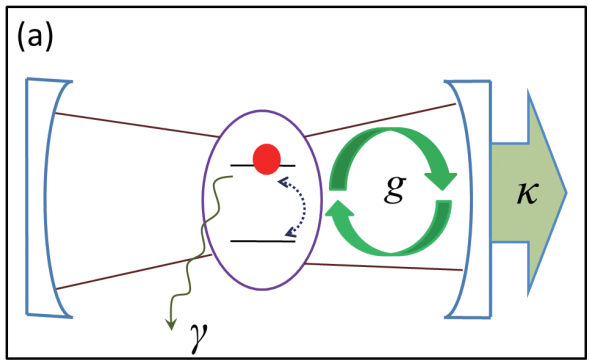}
\caption{(Color online) Schematic diagram of a two-level system or qubit
with dissipation rate $\protect\gamma ,$ which is coupled to a cavity with
loss rate $\protect\kappa $ by the qubit-cavity coupling strength $g.$}
\end{figure}
The Hamiltonian can be written (throughout this paper, we choose $\hbar =1$)
\cite{pra-74-033811} as%
\begin{eqnarray}
H &=&\frac{1}{2}\Delta \sigma _{z}+\sum_{k}\omega _{k,1}a_{k}^{\dagger
}a_{k}+\sum_{k}g_{k,1}(a_{k}^{\dagger }+a_{k})\sigma _{x}  \label{E1} \\
&&+\sum_{k}\omega _{k,2}b_{k}^{\dagger }b_{k}+\sum_{k}g_{k,2}(b_{k}^{\dagger
}+b_{k})\sigma _{x},  \notag
\end{eqnarray}%
where $\sigma _{x}=\sigma _{+}+\sigma _{-}$ and $i=1,2$, denote the
intrinsic and cavity baths of the qubit, respectively.

The baths experienced by the qubit can be characterized by a spectral
density $J_{i}(\omega )=\sum_{k}g_{k,i}^{2}\delta (\omega -\omega _{k,i}).$
To deal with the anti-rotating terms in Eq.~(\ref{E1}), we will perform a
unitary transformation on the Hamiltonian. This unitary transformation is
applied to the Hamiltonian $H$ as follows:
\begin{equation}
H^{\prime }=\exp (S)H\exp (-S),
\end{equation}%
with $S=S_{1}+S_{2},$ where $S_{i}$ ($i=1,2$) is given by
\begin{eqnarray}
S_{1} &=&\sum_{k}\frac{g_{k,1}}{\omega _{k,1}}\;\xi _{k,1}\left(
a_{k}^{\dagger }-a_{k}\right) \sigma _{x},  \label{EE3} \\
S_{2} &=&\sum_{k}\frac{g_{k,2}}{\omega _{k,2}}\;\xi _{k,2}\left(
b_{k}^{\dagger }-b_{k}\right) \sigma _{x}.  \label{EE4}
\end{eqnarray}%
In Eq.~(\ref{EE3}, \ref{EE4}), the parameter $\xi _{k,j}=\omega
_{k,j}/(\omega _{k,j}+\eta _{j}\,\Delta )$ is a $k$-dependent variable. Up
to order $g_{k,i}$, the transformed Hamiltonian $H^{(1)}$ can be written as%
\begin{eqnarray}
H^{\prime } &\approx &\frac{1}{2}\eta \,\Delta \sigma _{z}+\sum_{k}\omega
_{k,1}a_{k}^{\dagger }a_{k}+\sum_{k}\omega _{k,2}b_{k}^{\dagger }b_{k}
\notag \\
&&+\sum_{k}\tilde{g}_{k,1}\left( a_{k}^{\dagger }\sigma _{-}+a_{k}\sigma
_{+}\right)  \notag \\
&&+\sum_{k}\tilde{g}_{k,2}\left( b_{k}^{\dagger }\sigma _{-}+b_{k}\sigma
_{+}\right) ,  \label{e7}
\end{eqnarray}%
where

\begin{equation}
\tilde{g}_{k,i}=\left( \frac{2\eta _{i}\,\Delta }{\omega _{k,i}+\eta
_{i}\,\Delta }\right) g_{k,i},
\end{equation}

\begin{equation}
\eta =\eta _{1}\,\eta _{2},
\end{equation}%
\begin{equation}
\eta _{i}=\exp \left( -\sum_{k}\frac{2g_{k,i}^{2}{}}{\omega _{k,i}^{2}}\;\xi
_{k,i}^{2}{}\right) ,  \label{et2}
\end{equation}%
and $\eta =\eta _{1}\eta _{2}.$ Now the transformed Hamiltonian (\ref{e7})
has the same form as the Hamiltonian under the RWA, but its parameters have
been renormalized to include the effects of the anti-rotating terms related
with the intrinsic and cavity baths of the qubit. From the transformed
Hamiltonian $H^{\prime },$ one can see that, based on energy conservation,
the ground state of the transformed Hamiltonian $H^{\prime }$ is%
\begin{equation}
\left\vert g^{\prime }\right\rangle =\left\vert \downarrow \right\rangle
\otimes \prod\limits_{k}\left\vert 0_{k,1},0_{k,2}\right\rangle ,
\end{equation}
(using $\sigma _{z}\left\vert \downarrow \right\rangle =-\left\vert
\downarrow \right\rangle $) and the corresponding ground-state energy is $%
-\eta \;\Delta /2$. Therefore, the ground state of the original Hamiltonian $%
H$ is given by%
\begin{equation}
\left\vert g\right\rangle =\exp \left( -S\right) \left\vert g^{\prime
}\right\rangle ,
\end{equation}%
which is the dressed state of the qubit and the environment due to the
anti-rotating terms \cite{prl-103-147003,pra-80-053810}.

A qubit can experience different types of intrinsic baths. The most
commonly-used bath is the photon or phonon bath, which can be described by
an Ohmic spectrum. However, for many solid-state qubits (e.g.,
superconducting qubits), the dominant dissipation can be due to two-level
fluctuators, which behave like a low-frequency bath \cite{addie}. Here we
consider either an Ohmic or a low-frequency intrinsic bath. The Ohmic bath
with Drude cutoff is given by%
\begin{equation}
J_{1}^{\mathrm{Ohm}}(\omega )=\frac{2\alpha _{\mathrm{Ohm}}\omega }{1+\left(
\omega /\omega _{\mathrm{Ohm}}\right) ^{2}},
\end{equation}%
where $\omega _{\mathrm{Ohm}}$ is the high-frequency cutoff and $\alpha _{%
\mathrm{Ohm}}$ a dimensionless parameter characterizing the coupling
strength between the qubit and its intrinsic bath. Here the low-frequency
bath is written as
\begin{equation}
J_{1}^{\mathrm{low}}(\omega )=\frac{2\alpha _{\mathrm{low}}\omega }{\left(
\omega /\Delta \right) ^{2}+\left( \omega _{\mathrm{low}}/\Delta \right) ^{2}%
},  \label{E2}
\end{equation}%
where $\omega _{\mathrm{low}}$ is a characteristic frequency lower than the
qubit energy spacing $\Delta $ of the qubit, and $\alpha _{\mathrm{low}}$ is
a dimensionless coupling strength between the qubit and its intrinsic bath.
If $\omega \geq \omega _{\mathrm{low}}$, $J_{1}^{\mathrm{low}}(\omega )\sim
1/\omega $, corresponding to $1/f$ noise.

For a lossy cavity, the bath can be described by a Lorentzian spectral
density with a central frequency \cite{pra-79-042302}:
\begin{equation}
J_{2}(\omega )=\frac{g^{2}\lambda }{\pi \lbrack (\omega -\omega _{\mathrm{cav%
}})^{2}+\lambda ^{2}]},  \label{E3}
\end{equation}%
which corresponds to a single-mode cavity, with its frequency broadened by
the cavity loss. In Eq.~(\ref{E3}), $\lambda $ is the frequency width of the
cavity bath density spectrum, $\omega _{\mathrm{cav}}$ is the central
frequency of the cavity mode, and $g$ denotes the coupling strength between
the qubit and the cavity. Also, the parameter $\lambda $ is related to the
cavity bath correlation time and $\omega _{\mathrm{cav}}/\lambda $ is the
quality factor $Q$ of the cavity.

Using $J_{i}(\omega )=\sum g_{k,i}^{2}\delta (\omega -\omega _{k,i}),$ one
can derive from Eq.~(\ref{et2}) that $\eta _{i}$ is determined
self-consistently by the equation%
\begin{equation}
\log \eta _{i}+\int\limits_{0}^{\infty }\ \frac{2J_{i}(\omega )d\omega }{%
(\omega +\eta _{i}\,\Delta )^{2}}=0.
\end{equation}%
Below we will solve the equation of motion for the density matrix in
Hamiltonian (\ref{E1}) and obtain the qubit SE spectrum.

\subsection{Equation of Motion for the Density Matrix}

The equation of motion of the density matrix $\rho _{SB}$ for the whole
system, i.e., qubit system ($S$) and bath ($B$) is given by
\begin{equation}
\frac{d}{dt}\rho _{SB}(t)=-i[H,\rho _{SB}(t)].
\end{equation}%
After the unitary transformation (2), we have
\begin{equation}
\frac{d}{dt}\rho _{SB}^{\prime }(t)=-i[H^{\prime },\rho _{SB}^{\prime }(t)],
\label{ee7}
\end{equation}%
where $\rho _{SB}^{\prime }=\exp (S)\rho _{SB}\exp (-S)$ is the density
matrix of the whole system in the Schr{\"{o}}dinger picture with the
transformed Hamiltonian $H^{\prime }$ [i.e., Eq.~(\ref{e7})]. Below we solve
the transformed equation of motion, Eq.~(\ref{ee7}), in the interaction
picture with $H_{0}=\eta \,\Delta \sigma _{z}/2+\sum_{k,i}\omega
_{k,i}a_{k,i}^{\dagger }a_{k,i}$. In this interaction picture, the
transformed Hamiltonian $H^{\prime }$ can be written as%
\begin{equation}
V_{I}^{\prime }(t)=\sum_{k,i}\tilde{g}_{k,i}a_{k,i}^{\dagger }\sigma
_{-}\exp \left[ i(\omega _{k,i}-\eta \,\Delta )t\right] +\mathrm{H.c}.
\end{equation}%
The equation of motion for the density matrix $\rho _{SB}^{\prime I}(t)$ of
the whole system ($S$+$B$) can be written as
\begin{equation}
\frac{d}{dt}\rho _{SB}^{\prime I}(t)=-i[V_{I}^{\prime }(t),\rho
_{SB}^{\prime I}(t)].  \label{E-10}
\end{equation}%
with $\rho _{SB}^{\prime I}(t)=\exp \left( iH_{0}t\right) \rho _{SB}^{\prime
}(t)\exp \left( -iH_{0}t\right) .$

Integrating Eq.~(\ref{E-10}), we have%
\begin{equation}
\rho _{SB}^{\prime I}(t)=\rho _{SB}^{\prime
I}(t_{i})-i\int\limits_{t_{0}}^{t}[V_{I}^{\prime }(t^{\prime }),\rho
_{SB}^{\prime I}(t^{\prime })]dt^{\prime },  \label{E-11}
\end{equation}%
where $t_{0}$ is the initial time for the qubit-environment interaction to
turn on. Here we choose $t_{0}=0.$ Substituting $\rho _{SB}^{\prime I}(t)$
into Eq.~(\ref{E-10}), we obtain the equation of motion as%
\begin{eqnarray}
\frac{d}{dt}\rho _{SB}^{\prime I}(t) &=&-i[V_{I}^{\prime }(t),\rho
_{SB}^{\prime }(0)]  \label{E29} \\
&&-\int\limits_{0}^{t}[V_{I}^{\prime }(t),[V_{I}^{\prime }(t^{\prime }),\rho
_{SB}^{\prime I}(t^{\prime })]]dt^{\prime }.  \notag
\end{eqnarray}%
Using the Born approximation \cite{F1}, the density matrix $\rho
_{SB}^{\prime I}(t)$ in Eq.~(\ref{E29}) can be approximated by $\rho
_{SB}^{\prime I}(t)=\rho _{S}^{\prime I}(t)\rho _{B}(0).$ Tracing over the
degrees of freedom of the two baths, one obtain%
\begin{eqnarray}
\frac{d}{dt}\rho _{S}^{\prime I}(t) &=&-i\mathrm{Tr}_{R}[V_{I}^{\prime
}(t),\rho _{S}^{\prime }(0)\otimes \rho _{B}^{\prime }(0)]  \label{E30} \\
&&-\mathrm{Tr}_{R}\int\limits_{0}^{t}[V_{I}^{\prime }(t),[V_{I}^{\prime
}(t^{\prime }),\rho _{S}^{\prime I}(t^{\prime })\otimes \rho
_{B}(0)]]dt^{\prime },  \notag
\end{eqnarray}%
where $\rho _{S}^{\prime I}(t)=\mathrm{Tr}_{R}\left[ \rho _{SB}^{\prime I}(t)%
\right] .$ Because the two bosonic baths are assumed to be in thermal
equilibrium, it follows from the Bose-Einstein distribution that%
\begin{eqnarray}
\mathrm{Tr}_{R}\left( a_{k,i}^{\dagger }a_{k,i}\rho _{B}\right)  &=&n_{k,i},
\\
\mathrm{Tr}_{B}\left( a_{k,i}a_{k,i}^{\dagger }\rho _{B}\right)
&=&n_{k,i}+1,
\end{eqnarray}%
where $n_{k,i}$ is the thermal average boson number at mode $k$. Then,
substituting $V_{I}^{\prime }(t)$ into Eq.~(\ref{E30}), we have%
\begin{eqnarray}
&&\frac{d}{dt}\rho _{S}^{\prime I}(t)  \label{E34} \\
&=&-\sum_{k,i}\tilde{g}_{k,i}^{2}\int\limits_{0}^{t}f\!\left( t^{\prime
}\right) \exp \left[ i(\omega _{k,i}-\eta \,\Delta )(t-t^{\prime })\right]
dt^{\prime }-\mathrm{H.c.}  \notag
\end{eqnarray}%
where%
\begin{eqnarray}
f\left( t^{\prime }\right) \! &\!=\!&\!n_{k,i}\left[ \sigma _{-}\sigma
_{+}\rho _{S}^{\prime I}\left( t^{\prime }\right) -\sigma _{+}\rho
_{S}^{\prime I}(t^{\prime })\sigma _{-}\right]  \\
&&\!+\left( n_{k,i}+1\right) \left[ \rho _{S}^{\prime I}(t^{\prime })\sigma
_{+}\sigma _{-}-\sigma _{-}\rho _{S}^{\prime I}(t^{\prime })\sigma _{+}%
\right] .  \notag
\end{eqnarray}%
On the right-hand side of Eq.~(\ref{E34}), the terms related with $n_{k,i}$
and $n_{k,i}+1$ describe, respectively, the decay and excitation processes,
with the rates depending on the temperature. Here, for simplicity, we study
the zero-temperature case with $n_{k,i}=0,$ i.e., only the spontaneous decay
occurs, which corresponds to a purely dissipative process. The appendix
gives the solution of Eq.~(\ref{E34}) for the reduced density matrix $\rho
_{S}^{\prime I}(t)$ of the qubit in the interaction picture. With the
solution for $\rho _{S}^{\prime I}(t),$ one can derive the reduced density
matrix $\rho _{S}^{\prime }(t)$ of the qubit in the Schr{\"{o}}dinger
picture:%
\begin{widetext}
\begin{eqnarray}
\rho _{S}^{\prime }(t) &=&\exp \left( i\eta \,\Delta \sigma _{z}t/2\right)
\rho _{S}^{\prime I}(t)\exp \left( i\eta \,\Delta \sigma _{z}t/2\right)
\label{E27} \\
&=&\left(
\begin{array}{cc}
\mathcal{L}^{-1}\left[ \frac{\rho _{22}^{\prime }(0)}{p+A_{+}+A_{-}}\right]
& \mathcal{L}^{-1}\left[ \frac{\rho _{21}^{\prime }(0)}{p+A_{+}}\right]
\mathrm{e}^{-it\text{$\Delta $}\text{$\eta $}} \\
\mathcal{L}^{-1}\left[ \frac{\rho _{12}^{\prime }(0)}{p+A_{-}}\right]
\mathrm{e}^{it\text{$\Delta $}\text{$\eta $}} & \mathcal{L}^{-1}\left[ \frac{%
\rho _{_{22}}^{\prime }(0)}{p}-\frac{\rho _{_{22}}^{\prime }(0)}{%
p+A_{+}+A_{-}}+\frac{\rho _{_{11}}^{\prime }(0)}{p}\right]
\end{array}%
\right) .  \notag
\end{eqnarray}
\end{widetext}Because the reduced density matrix $\rho _{S}(t)$ in the Schr{%
\"{o}}dinger picture with the original Hamiltonian $H$ [i.e. Eq.~(\ref{E1})]
is related to $\rho _{S}^{^{\prime }}(t)$ by the relation
\begin{equation}
\rho _{S}(t)=\mathrm{Tr}_{B}\left[ \exp \left( -S\right) \rho _{S}^{\prime
}(t)\rho _{B}\exp \left( S\right) \right] .
\end{equation}%
Then, using $\exp \left( S\right) =\cosh Y+\sigma _{x}\sinh Y$, with $%
Y=\sum_{k,i}g_{k,i}\xi _{k,i}(a_{k,i}^{\dagger }-a_{k,i})/\omega _{k,i}$,
and tracing over the degrees of freedom of the two baths, we obtain%
\begin{equation}
\rho _{S}(t)=\frac{1+\eta }{2}\rho _{S}^{\prime }(t)+\frac{1-\eta }{2}\sigma
_{x}\,\rho _{S}^{\prime }(t)\,\sigma _{x}.
\end{equation}

\subsection{Derivation of the spontaneous Emission Spectrum}

When measured by an ideal system with negligible bandwidth, the spontaneous
emission spectrum can be given by \cite{hjc}%
\begin{equation}
P(\omega )\propto \int\limits_{0}^{\infty }\!\!dt\int\limits_{0}^{\infty
}\!\!dt^{\prime }\exp \left[ -i\omega (t-t^{\prime })\right] C(t,t^{\prime
}),
\end{equation}%
with the two-time correlation function
\begin{eqnarray}
C(t,t^{\prime }) &=&\left\langle \sigma _{+}(t)\sigma _{-}(t^{\prime
})\right\rangle  \\
&=&\left\langle \psi (0)\right\vert \sigma _{+}(t)\sigma _{-}(t^{\prime
})\left\vert \psi (0)\right\rangle ,  \notag
\end{eqnarray}%
where the $\sigma _{\pm }(t)=\exp \left( iHt\right) \sigma _{\pm }\exp
\left( -iHt\right) $ are the raising ($+$) and lowering ($-$) Pauli matrices
in the Heisenberg picture and $\left\vert \psi (0)\right\rangle $ is the
initial state of the whole system, which remains unchanged in the Heisenberg
picture. Using $\left\vert \psi ^{\prime }(0)\right\rangle =\exp \left(
S\right) \left\vert \psi (0)\right\rangle ,$ the two-time correlation
function can be written as%
\begin{equation}
C(t,t^{\prime })=\left\langle \psi ^{\prime }(0)\right\vert \mathrm{e}%
^{iH^{\prime }t}\sigma _{+}^{\prime }\mathrm{e}^{-iH^{\prime }t}\mathrm{e}%
^{iH^{\prime }t^{\prime }}\sigma _{-}^{\prime }\mathrm{e}^{-iH^{\prime
}t^{\prime }}\left\vert \psi ^{\prime }(0)\right\rangle ,
\end{equation}%
where $\sigma _{\pm }^{\prime }=\exp \left( S\right) \sigma _{\pm }\exp
\left( -S\right) $. Because the zero-temperature case is considered here,
this two-time correlation function can be approximated as%
\begin{equation}
C(t,t^{\prime })\approx \left\langle \psi ^{\prime }(0)\right\vert \mathrm{e}%
^{iH^{\prime }t}\tilde{\sigma}_{+}^{\prime }\mathrm{e}^{-iH^{\prime }t}%
\mathrm{e}^{iH^{\prime }t^{\prime }}\tilde{\sigma}_{-}^{\prime }\mathrm{e}%
^{-iH^{\prime }t^{\prime }}\left\vert \psi ^{\prime }(0)\right\rangle ,
\end{equation}%
with%
\begin{eqnarray}
\tilde{\sigma}_{\pm }^{\prime } &=&\prod\limits_{k}\left\langle 0_{k^{\prime
},1},0_{k^{\prime },2}\right\vert \sigma _{\pm }^{\prime }\left\vert
0_{k,1},0_{k,2}\right\rangle   \notag \\
&=&\frac{1+\eta }{2}\;\sigma _{\pm }+\frac{1-\eta }{2}\;\sigma _{\mp }.
\label{EE30}
\end{eqnarray}%
In deriving Eq.~(\ref{EE30}), the two baths are assumed in the ground state
for the zero-temperature case. Therefore, from Eq.~(\ref{EE30}), we have%
\begin{eqnarray}
&&C(t,t^{\prime })\approx \left( \frac{1+\eta }{2}\right) ^{2}\left\langle
\sigma _{+}\left( t\right) \sigma _{-}\left( t^{\prime }\right)
\right\rangle _{H^{\prime }}  \notag \\
&&+\frac{1-\eta ^{2}}{4}\left( \left\langle \sigma _{+}\left( t\right)
\sigma _{+}\left( t^{\prime }\right) \right\rangle _{H^{\prime
}}+\left\langle \sigma _{-}\left( t\right) \sigma _{-}\left( t^{\prime
}\right) \right\rangle _{H^{\prime }}\right)   \notag \\
&&+\left( \frac{1-\eta }{2}\right) ^{2}\left\langle \sigma _{-}\left(
t\right) \sigma _{+}\left( t^{\prime }\right) \right\rangle _{H^{\prime }},
\label{70}
\end{eqnarray}%
where the expectation value of the operator in the transformed Hamiltonian $%
H^{\prime }$ is%
\begin{eqnarray}
&&\left\langle \sigma _{\alpha }\left( t\right) \sigma _{\beta }\left(
t^{\prime }\right) \right\rangle _{H^{\prime }} \\
&=&\left\langle \psi ^{\prime }(0)\right\vert \mathrm{e}^{iH^{\prime
}t}\sigma _{\alpha }^{\prime }\mathrm{e}^{-iH^{\prime }t}\mathrm{e}%
^{iH^{\prime }t^{\prime }}\sigma _{\beta }^{\prime }\mathrm{e}^{-iH^{\prime
}t^{\prime }}\left\vert \psi ^{\prime }(0)\right\rangle ,  \notag
\end{eqnarray}%
with $\alpha ,\beta =\pm .$

In our case, $\eta $ is very close to $1$. Therefore, Eq.~(\ref{70}) can be
further approximated as%
\begin{equation}
C(t,t^{\prime })\approx C(t,t^{\prime })_{H^{\prime }},
\end{equation}%
with $C(t,t^{\prime })_{H^{\prime }}=$ $\left\langle \sigma _{+}\left(
t\right) \sigma _{-}\left( t^{\prime }\right) \right\rangle _{H^{\prime }}.$
For a qubit state specified by a density matrix $\rho (t),$ we can formulate
the expectation values of $\sigma _{+}(t),$ $\sigma _{-}(t)$ and $\sigma
_{+}(t)\sigma _{-}(t)$ by the matrix elements, $\left\langle \sigma
_{+}(t)\right\rangle =\left\langle \sigma _{-}(t)\right\rangle ^{\ast }=\rho
_{21}(t)$ and $C(t,t)=\rho _{11}(t).$ According to the quantum regression
theorem \cite{hjc}, the correlation function becomes%
\begin{eqnarray}
&&C(t,t+\tau )_{H^{\prime }} \\
&=&\mathcal{L}^{-1}\left( \frac{1}{p+A_{+}}\right) _{\!\tau }\mathrm{e}%
^{-i\Delta \eta \text{$\tau $}}\rho _{11}^{\prime }(t).  \notag
\end{eqnarray}%
In a spontaneous emission process, the initial state is an excited state $%
\left\vert \psi (0)\right\rangle =\exp \left( -S\right) \left\vert \uparrow
\right\rangle \otimes \prod\limits_{k}\left\vert
0_{k,1},0_{k,2}\right\rangle $, which can be achieved by $\sigma
_{x}\left\vert g\right\rangle .$ Then, the initial state in the transformed
Hamiltonian (\ref{e7}) is $\left\vert \psi ^{\prime }(0)\right\rangle
=\left\vert \uparrow \right\rangle \otimes \prod\limits_{k}\left\vert
0_{k,1},0_{k,2}\right\rangle ,$ i.e. $\rho _{11}^{\prime }=1.$ Therefore,
from Eq.~(\ref{A-12}) in the Appendix A, the dynamics evolution of $\rho
_{11}^{\prime }$ can be expressed as%
\begin{equation}
\rho _{11}^{\prime }(t)=\mathcal{L}^{-1}\left( \frac{1}{p+A_{+}+A_{-}}%
\right) _{\!t}.
\end{equation}%
Using the Schr{\"{o}}dinger equation (see Appendix B) \cite{pra77}, we have%
\begin{eqnarray}
&&\mathcal{L}^{-1}\left( \frac{1}{p+A_{+}+A_{-}}\right) _{\!t} \\
&=&\mathcal{L}^{-1}\left( \frac{1}{p+A_{+}}\right) _{\!t}\times \mathcal{L}%
^{-1}\left( \frac{1}{p+A_{-}}\right) _{\!t},  \notag
\end{eqnarray}%
where $\mathcal{L}^{-1}\left( \frac{1}{p+A_{+}}\right) $ and $\mathcal{L}%
^{-1}\left( \frac{1}{p+A_{-}}\right) $ are conjugate quantities (see
Appendix A). From%
\begin{eqnarray}
&&C(t,t+\tau )_{H^{\prime }} \\
&=&\mathrm{e}^{-i\text{$\Delta $}\text{$\eta $}\left( \text{$\tau +t$}%
\right) }\mathcal{L}^{-1}\left( \frac{1}{p+A_{+}}\right) _{\!\left( \tau
+t\right) }\!\times \mathcal{L}^{-1}\left( \frac{1}{p+A_{-}}\right)
_{\!t}e^{i\text{$\Delta $}\text{$\eta $}t},  \notag
\end{eqnarray}%
we obtain the two-time correlation function for any $t$ and $t^{\prime }$,%
\begin{eqnarray}
C(t,t^{\prime })_{H^{\prime }} &=&\mathcal{L}^{-1}\left( \frac{1}{p+A_{+}}%
\right) _{\!t}\mathrm{e}^{-i\text{$\Delta $}\text{$\eta $}t} \\
&&\times \mathcal{L}^{-1}\left( \frac{1}{p+A_{-}}\right) _{\!t^{\prime }}%
\mathrm{e}^{-i\text{$\Delta $}\text{$\eta $}t^{\prime }}.  \notag
\end{eqnarray}%
\ Finally, using the Wiener-Khinchin theorem, the SE spectrum is given by%
\begin{eqnarray}
P(\omega ) &\propto &\int\limits_{0}^{\infty }\!\!dt\int\limits_{0}^{\infty
}\!\!dt^{^{\prime }}\exp \left[ -i\omega (t-t^{^{\prime }})\right]
C(t,t^{^{\prime }}) \\
&=&\left\vert \mathcal{F}\left[ \mathcal{L}^{-1}\left( \frac{1}{p+A_{+}}%
\right) \mathrm{e}^{-i\text{$\Delta $}\text{$\eta \tau $}}\right]
\right\vert ^{2}  \notag \\
&=&\frac{1}{\left[ \omega -\Delta \eta -R\left( \omega \right) \right]
^{2}+\Gamma \left( \omega \right) ^{2}}.  \notag
\end{eqnarray}

\section{Dependence of the spontaneous emission spectrum on the baths}

We will show the SE spectrum of the qubit in resonance with the cavity
central frequency ($\Delta =\omega _{\mathrm{cav}}$) as a function of the
microwave probe frequency for three cases: weak, strong and ultra-strong
qubit-cavity couplings.

\textit{Weak coupling} means the qubit-cavity coupling strength $g$ is less
than the sum of the dissipation rate of the qubit and the cavity. The
dissipation rates of the qubit due to its intrinsic bath is approximately
denoted as $\Gamma _{\mathrm{qb}},$ which can be $\Delta \: \alpha _{\mathrm{%
Ohm}}$ or $\Delta \: \alpha _{\mathrm{low}}.$ And the dissipation rate due
to the cavity bath can be approximately as the spectrum width of the cavity
spectral density $\lambda .$ So$\ $weak coupling is express as $g<\left(
\Gamma _{\mathrm{qb}}+\lambda \right) .$

\textit{Strong coupling} means that the qubit-cavity coupling strength $g$
is larger than the sum of the dissipation rate of the qubit and the cavity: $%
g>\left( \Gamma _{\mathrm{qb}}+\lambda \right) $, but it is typically two
orders of magnitude smaller than the qubit energy spacing $\Delta $ and the
cavity frequency $\omega _{\mathrm{cav}}$, i.e., $g\sim 10^{-2}\Delta ,$
such as the case in Ref.~[\onlinecite{nature-431-162}].

\textit{Ultra-strong coupling} means that the qubit-cavity coupling $g$ is
to a significant fraction of the transition frequency $\Delta $ (e.g., $%
g\gtrsim 0.1\Delta $). This case extends to the fine-structure limit for the
maximal value of an electric-dipole coupling.

The energy spectral density of a bath plays an important role in determining
the energy-shift direction and the asymmetry of the SE spectrum. In Fig.~2,
we show the spectral densities for both intrinsic and cavity baths of the
qubit. For the intrinsic bath of the qubit, both a low-frequency bath and an
Ohmic bath are considered. The spectral density of the cavity bath is
symmetric about the central frequency $\omega _{_{\mathrm{cav}}}$ of the
cavity.
\begin{figure}[tbp]
\includegraphics[width=7cm,clip]{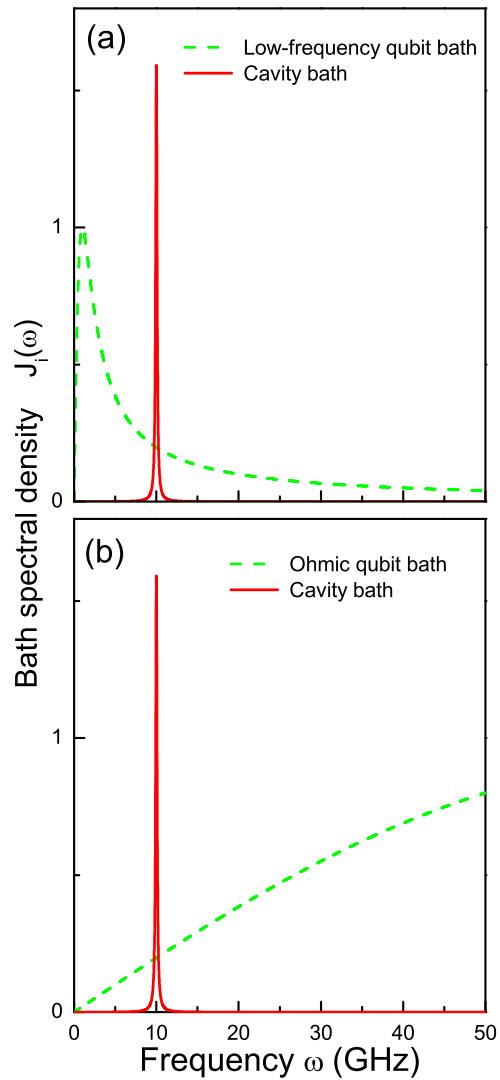}
\caption{(Color online) The spectrum density of the qubit environment. (a)
Lorentzian cavity bath and low-frequency intrinsic bath of the qubit. (b)
Lorentzian cavity bath and Ohmic intrinsic bath of the qubit. From Fig.~(a)
and (b), we see that the dominant regimes of the low-frequency and Ohmic
qubit bath spectral density are different.}
\end{figure}

Considering the experimental parameters \cite{nature-431-162,add9}, we
assume the qubit energy spacing to be$\ \Delta =10$ $\mathrm{GHz}$. The
dimensionless coupling strength $\alpha $ between the qubit and its
intrinsic bath (either Ohmic- or low-frequency bath) is fixed at $\alpha
=10^{-4},$ which implies that the decay rate of the intrinsic bath is $%
\Gamma _{\mathrm{qb}}\sim 1$ $\mathrm{MHz}.$

\subsection{Effect of the cavity bath on the spontaneous emission spectrum}

To illustrate the effect of the cavity bath with a symmetric spectral
density, in Fig.~3 we show the qubit SE spectrum when only the cavity bath
is present. To enhance these features further, we choose a low quality
factor $Q=10^{2}$, and plot the SE spectra in Fig.~3(a) for strong ($%
g=10^{2} $ $\mathrm{MHz}$) and ultrastrong ($g=$ $10^{3},$ $2\times 10^{3}$ $%
\mathrm{MHz}$) qubit-cavity coupling. Due to the scope of application of
this method, only the lower bound of the ultra-strong coupling $g=$ $%
0.1\Delta $ and $g=0.2\Delta $ are considered. As a comparison, the results
under the RWA are given in Fig.~3(b). Also, the SE spectra without and with
RWA for $Q=10^{3}$ are plotted in Fig.~3(c,d). In the case of strong
qubit-cavity coupling, the two peaks of the vacuum Rabi splitting are nearly
symmetric about $\omega =\Delta $, almost coinciding with the results under
the RWA. When the qubit-cavity coupling increases, the height and the
position asymmetry (about qubit energy spacing $\Delta $) of the two peaks \
becomes more apparent, in sharp contrast to the symmetric SE peaks obtained
under RWA [see Fig.~3(a,b) and (c,d)]. As we know, if the RWA is used, the
qubit-cavity coupling term $\sum\limits_{k}g_{k,2}(a_{k,2}^{\dagger
}+a_{k,2})(\sigma _{+}+\sigma _{-})$ in the Hamiltonian $H$ becomes $%
\sum\limits_{k}g_{k,2}(a_{k,2}^{\dagger }\sigma _{-}+a_{k,2}\sigma _{+})$.
The energy spectral density in the regions lower and higher than the central
frequency of the cavity (related to absorbing and emitting a single photon
in the cavity) are identical. Therefore, when the qubit energy spacing is in
resonance with the cavity central frequency, the coupling strength is
symmetric about the qubit energy spacing for the absorption and emission
processes.

While taking into account the anti-rotating terms, the coupling term
becomes\ $\sum\limits_{k}\tilde{g}_{k,2}(a_{k,2}^{\dagger }\sigma
_{-}+a_{k,2}\sigma _{+})$ in the transformed Hamiltonian $H^{\prime }$, with
a renormalized coupling strength $\tilde{g}_{k,2}=2\eta _{2}\,\Delta
g_{k,2}/\left( \omega _{k,2}+\eta _{2}\,\Delta \right) $ . Obviously,
\textit{the renormalized coupling strength }$\tilde{g}_{k,2}$\textit{\
induces the spectral asymmetry}: for a symmetric spectral density of the
cavity bath , in the region $\omega _{k,2}<\Delta $, due to $2\eta
_{2}\,\Delta /\left( \omega _{k,2}+\eta _{2}\,\Delta \right) >1,$ the
renormalized interaction $\tilde{g}_{k,2}$ is \textit{larger} than $g_{k,2}$%
. However, in the region $\omega _{k,2}>\Delta ,$ owing to $2\eta
_{2}\,\Delta /\left( \omega _{k,2}+\eta _{2}\,\Delta \right) <1,$ the
effective coupling strength $\tilde{g}_{k,2}$ is \textit{smaller} than $%
g_{k,2}$.

These results (with and without the RWA) indicate that the RWA cannot be
used in the range of ultrastrong qubit-cavity coupling. The general tendency
observed here is that \textit{the RWA overestimates the frequency shift in
the low-energy regime} $\omega \sim -g$, \textit{while it under-estimates
the frequency shift in the high-energy regime} $\omega \sim g.$ Our results
are consistent with the results in Ref.~[\onlinecite{pra-80-033846}].
\begin{figure}[tbp]
\includegraphics[width=7cm,clip]{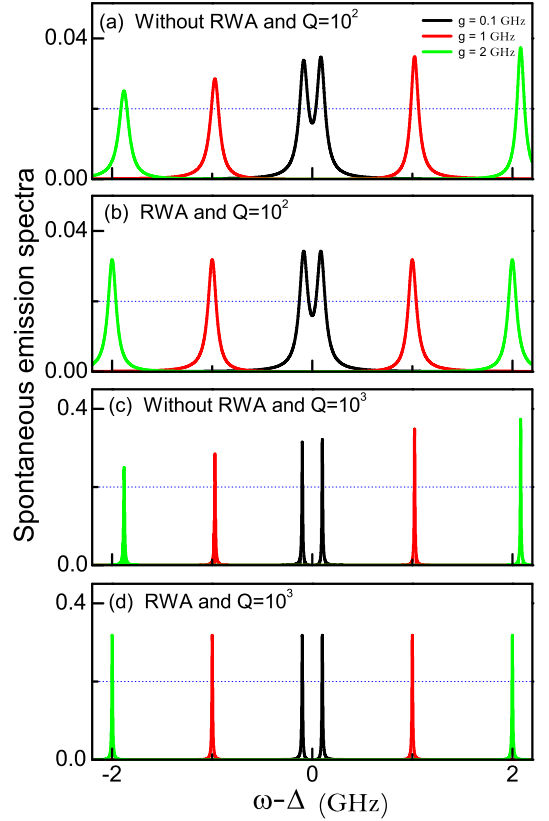}
\caption{(Color online) Spontaneous emission spectra of the qubit only in
the cavity bath (symmetric spectral density). Figure (a) and (c) are the
results without RWA with $Q=10^{2}$ and $Q=10^{3}$. Figures (b) and (d) are
the results of the RWA with $Q=10^{2}$ and $Q=10^{3}$. To see the height
asymmetry of two peaks clearly, the horizontal grid lines are plotted as
reference. Note that in (a) and (c), the two peaks of spontaneous emission
spectrum present obvious height and position asymmetry (about $\protect%
\omega =\Delta $) in ultra-strong qubit-cavity coupling.}
\end{figure}

\subsection{Combined effect of both intrinsic and cavity baths on the
spontaneous emission spectrum}

\label{b}

\begin{figure}[tbp]
\includegraphics[width=7cm,clip]{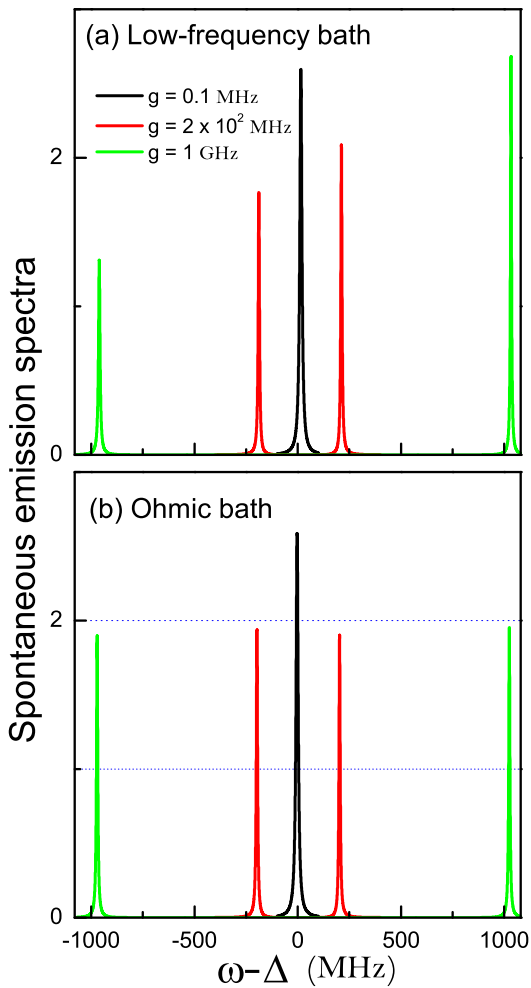}
\caption{(Color online) Spontaneous emission spectra of the qubit in
resonance with the central frequency $\protect\omega _{_{\mathrm{cav}}}$ of
the cavity for weak, strong and the lower bound of the ultra-strong
qubit-cavity interactions. The cavity quality factor is $Q=10^{4}$. (a)
coupling strength to the low-frequency intrinsic bath of the qubit $\protect%
\alpha _{\mathrm{low}}=10^{-4}$, (b) coupling strength to the Ohmic
intrinsic bath of the qubit $\protect\alpha _{\mathrm{Ohm}}=10^{-4}$. To see
the height asymmetry of the peaks clearly, the horizontal grid lines are
plotted as reference. Note that (a) demonstrates the obvious height
asymmetry in the case of strong and ultra-strong qubit-cavity coupling and
the asymmetry increases as the qubit-cavity coupling grows. Figure (b) shows
inverted height asymmetry of two peaks from (a) in the strong qubit-cavity
coupling case, but as the qubit-cavity coupling increases to the
ultra-strong regime, the height asymmetry of the right and left spectral
peaks are inverted.}
\end{figure}

\begin{figure}[tbp]
\includegraphics[width=7cm,clip]{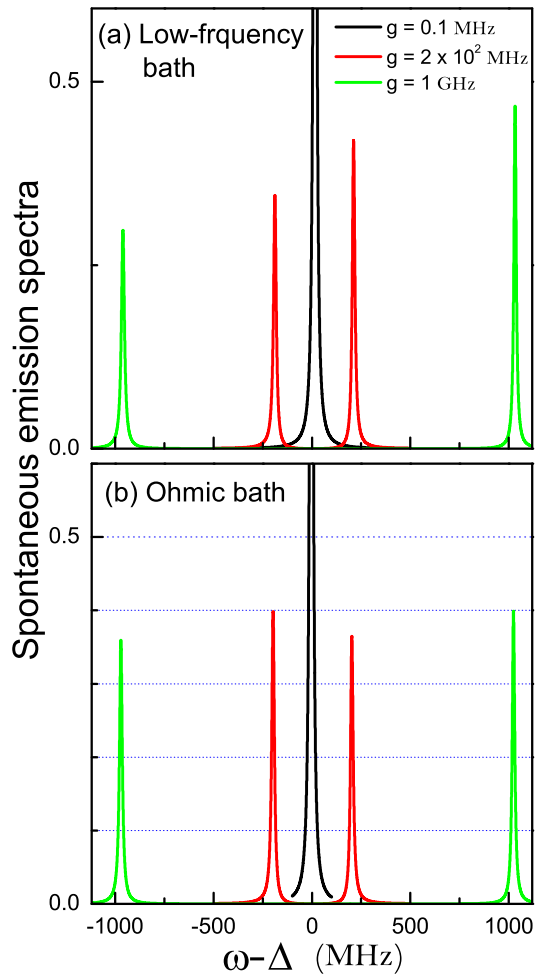}
\caption{(Color online) Spontaneous emission spectra of the qubit in
resonance with the central frequency $\protect\omega _{_{\mathrm{cav}}}$ of
cavity for weak, strong and the lower bound of the ultra-strong qubit-cavity
interactions. The cavity quality factor is $Q=10^{3}$. (a) coupling strength
to the low-frequency intrinsic bath of the qubit $\protect\alpha _{\mathrm{%
low}}=10^{-4}$, (b) coupling strength to the Ohmic intrinsic bath of the
qubit $\protect\alpha _{\mathrm{Ohm}}=10^{-4}$. To see clearly the height
asymmetry of the peaks, horizontal grid lines are plotted as reference. Note
that (a) demonstrates an obvious height asymmetry in the case of strong and
ultra-strong qubit-cavity coupling and the asymmetry increases as the
qubit-cavity coupling grows. Figure (b) shows the inverted height asymmetry
of two peaks from (a) in the strong qubit-cavity coupling case, but as the
qubit-cavity coupling increases to the ultra-strong regime, the height
asymmetry of the right and left spectral peaks are inverted.}
\end{figure}

Although a high-$Q$ seems plausible for minimizing the loss of the cavity,
it limits the measurement speed. Here, we consider a cavity with the quality
factor \cite{pra-69-062320,nature-431-162} $Q=10^{4}$ in the presence of a
qubit (see Fig.~4). We also plot the SE spectrum for $Q=$ $10^{3}$ in
Fig.~5,\ and see how the quality factor affects the results. The dissipation
rate of the cavity is approximately to the spectral width $\lambda $ of the
cavity bath. If $Q=10^{4},$ the dissipation rate of the cavity bath is about
$1$ $\mathrm{MHz}$ (the same order of magnitude of the bath dissipation
rate). Figures 4(a) and 4(b) show the spectra of the qubit coupled with a
low-frequency and an Ohmic intrinsic baths, respectively. From Fig.~4(a), we
see that in the case of weak qubit-cavity coupling $g=10^{-5}\Delta =0.1$ $%
\mathrm{MHz}$, the SE spectrum is a single peak with the central frequency
larger than the energy spacing $\Delta $ of the bare qubit, which
corresponds to a blue-shift. In the case of \textit{strong} qubit-cavity
coupling, $g=2\times 10^{-2}\Delta =2\times 10^{2}$ $\mathrm{MHz}$, the SE
spectrum shows the vacuum Rabi splitting, with the two height asymmetric
peaks, just as shown in Ref.~[\onlinecite{nature-431-162}]. By further
increasing the qubit-cavity coupling $g=10^{-1}\Delta =1$ $\mathrm{GHz},$
not only the height of the SE peaks but also their positions demonstrate a
strong asymmetry about $\omega =\Delta $. The qubit-cavity coupling strength
$g=10^{-1}\Delta $ is the lower bound of the ultra-strong coupling regime.

Figure~4(b) shows the SE spectrum of a qubit in the Ohmic intrinsic bath.
For a weak qubit-cavity coupling, the central frequency of the SE spectrum
shifts to an energy slightly lower than the energy spacing $\Delta $ of the
bare qubit, which corresponds to a red-shift. This energy-shift direction is
opposite to the case when the\ qubit is in a low-frequency intrinsic bath
[see Fig.~4(a)], and the energy shift is smaller. For a strong qubit-cavity
coupling, the two SE peaks also show a weak asymmetry , with the left peak
higher than the right peak. This peak asymmetry is inverted from the same
case in Fig.~4(a). As the qubit-cavity coupling increases to the
ultra-strong regime, the height asymmetry of the left and right SE spectrum
are inverted [see Fig.~4(b)]. The SE spectra of the qubit in a cavity with
quality factor $Q=10^{3}$ are shown in Fig.~5, which present nearly the same
features as in Fig.~4, in addition to the broader SE peaks in Fig.~5. This
is because of an increased dissipation rate of the qubit induced by a larger
cavity dissipation rate. These results are briefly summarized in table I.

\begin{table*}[tbp]
\caption{Summary of our main results for the spontaneous emission spectra in
the case of weak, strong and ultra-strong qubit-cavity coupling. The spectra
with symmetric peaks (S), height asymmetric peaks (AS) and very asymmetric
peaks (VAS) are abbreviated as S, AS and VAS. The asterisks indicate which
are inverted peak height asymmetry from the AS in the same case of the
low-frequency bath. These results are described in detail in the main text. }%
\begin{tabular}{|c|c|c|c|c|}
\hline
\multirow{2}{*} {Bath} & Cavity & \multicolumn{3}{|c|}{qubit-cavity coupling}
\\
\hhline{~~---} & quality factor & weak & strong & ultra-strong \\
\hline\hline
\multirow{4}{*}{Low-frequency} & \multirow{2}{*}{high Q} & single & %
\multirow{2}{*}{AS} & \multirow{2}{*}{VAS} \\
\hhline{~~~~~} &  & peak &  &  \\
\hhline{~----} & \multirow{2}{*}{low Q} & single & \multirow{2}{*}{AS} & %
\multirow{2}{*}{VAS} \\
\hhline{~~~~~} &  & peak &  &  \\ \hline
\multirow{4}{*}{Ohmic} & \multirow{2}{*}{high Q} & single & %
\multirow{2}{*}{AS*} & \multirow{2}{*}{AS} \\
\hhline{~~~~~} &  & peak &  &  \\
\hhline{~----} & \multirow{2}{*}{low Q} & single & \multirow{2}{*}{AS*} & %
\multirow{2}{*}{AS} \\
\hhline{~~~~~} &  & peak &  &  \\ \hline
\end{tabular}%
\end{table*}

Figures 4 and 5 show that the two SE peaks of the qubit in both a
low-frequency and Ohmic intrinsic baths have very different behaviors; the
right SE peak is higher than the left SE peak in both strong and ultrastrong
qubit-cavity coupling regimes, while the right SE peak is lower than the
left SE peak in the strong qubit-cavity coupling regime, and in the lower
bound of the ultra-strong qubit-cavity coupling regime, the left SE peak is
slightly lower than the right peak. Therefore, the different asymmetric
behaviors of the SE spectrum of the qubit in the low-frequency and Ohmic
baths may be used to distinguish the intrinsic noise of the qubit. In the
experiment in Ref.~[\onlinecite{nature-431-162}], the height asymmetric SE
spectrum of the qubit in the strong qubit-cavity coupling regime shows that
the right SE peak is higher than the left SE peak. This reveals that \textit{%
the low-frequency intrinsic noise is dominant in the superconducting qubit
in Ref.~[\onlinecite{nature-431-162}]}.

\section{Discussion and conclusion}

Below we discuss the reason why the SE spectrum in the low-frequency
intrinsic bath is different from that in the Ohmic intrinsic bath. We begin
with the standard Born-Markov master equation (at zero temperature) \cite%
{hjc,Agarwal}, which is usually derived under the RWA \cite{louisell}%
\begin{eqnarray}
\dot{\rho} &=&-i\left[ H,\rho \right] +\gamma \left( 2\sigma _{-}\rho \sigma
_{+}-\sigma _{+}\sigma _{-}\rho -\rho \sigma _{+}\sigma _{-}\right)
\label{60} \\
&&+\kappa \left( 2a\rho a^{\dagger }-a^{\dagger }a\rho -\rho a^{\dagger
}a\right) .  \notag
\end{eqnarray}%
Equation~(\ref{60}) is typically used to describe the whole system
consisting of an atom and a (single mode) cavity with dissipation via two
channels: the dissipation of the qubit due to a free-space mode (the term
proportional to $\gamma $), and the dissipation of the qubit due to the
cavity loss (the term proportional to $\kappa $). In quantum optics, the SE
spectrum is often calculated by substituting the Jaynes--Cummings model
Hamiltonian $H_{\mathrm{I}}^{\mathrm{JC}}=g\left( \sigma _{-}a^{\dagger
}+\sigma _{+}a\right) $ as $H,$ into Eq.~(\ref{60}). Then the vacuum Rabi
frequency spitting with two symmetric SE peaks is derived, in the strong
qubit-cavity coupling, where the position of the two peaks are exactly at $%
\omega =\pm g$. This RWA result is different from the experimental result
\cite{nature-431-162}, where the two SE peaks are asymmetric about $\omega
=\Delta $ (the energy level spacing of the bare qubit). As shown in Sec. \ref%
{b}, the height asymmetry of the SE spectrum is obtained beyond RWA.
Therefore, it can be concluded that \textit{the anti-rotating terms produce
an asymmetric SE spectrum of the qubit}.

In conclusion, we discussed the SE spectrum of a qubit in the environment
described by two baths: intrinsic bath and cavity bath. We only consider
that the central frequency of the cavity mode is resonant with the qubit
energy spacing ($\omega _{_{\mathrm{cav}}}=\Delta $). We analyze in detail
the qubit's SE spectrum in the weak, strong, and the lower bound of the
ultra-strong coupling regimes, and compare the SE spectra in two kinds of
qubit baths: low-frequency and Ohmic baths. In the low-frequency bath, the
height asymmetry of the vacuum Rabi splitting peaks increases as the
coupling strength grows. However, for the Ohmic bath, the height asymmetry
of the SE spectrum is reduced, then the height asymmetry of the left and
right peaks are inverted, when the coupling strength is increased. All of
these results show that for strong qubit-cavity coupling, the asymmetry of
the splitting peaks in the SE spectrum comes from the anti-rotating terms
and the non-constant spectral density of the bath.

\vskip 0.5cm

{\noindent {\large \textbf{Acknowledgements}}}

We are very grateful to Adam Miranowicz for his very helpful comments. FN
acknowledges partial support from DARPA, the Laboratory of Physical
Sciences, National Security Agency, Army Research Office, National Science
Foundation grant No. 0726909, JSPS-RFBR contract No. 09-02-92114,
Grant-in-Aid for Scientific Research (S), MEXT Kakenhi on Quantum
Cybernetics, and Funding Program for Innovative R\&D on S\&T (FIRST). J. Q.
You acknowledges partial support from the National Natural Science
Foundation of China under Grant No. 10625416, the National Basic Research
Program of China under Grant No. 2009CB929300 and the ISTCP under Grant No.
2008DFA01930. X.-F. Cao acknowledges support from the National Natural
Science Foundation of China under Grant No. 10904126 and Fujian Province
Natural Science Foundation under Grant No. 2009J05014.

\bigskip

\appendix

\section*{Appendix A: Solution of the equation of motion of the density
matrix}

\label{sec:appendix}

\setcounter{equation}{0} \renewcommand{\theequation}{A.\arabic{equation}}

This appendix offers detailed calculations for solving the master equation
in Eq.~(\ref{E34}). We use the basis $|1\rangle =\left\vert \downarrow
\right\rangle $ and $|2\rangle =\left\vert \uparrow \right\rangle $, where $%
\sigma _{z}\left\vert \downarrow \right\rangle =\left\vert \downarrow
\right\rangle ,$ $\sigma _{z}\left\vert \uparrow \right\rangle =-\left\vert
\uparrow \right\rangle ,$ to define the reduced density matrices of the
qubit. By using the Laplace transform%
\begin{equation}
\tilde{\rho}(p)=\mathcal{L}\left[ \rho (t)\right] =\int\limits_{0}^{\infty
}dt\rho (t)\exp \left( -pt\right)
\end{equation}%
and the convolution theorem%
\begin{equation}
\mathcal{L}\left[ \int\limits_{0}^{t}dt^{\prime }f_{1}(t^{\prime
})f_{2}(t-t^{\prime })\right] =\mathcal{L}\left[ f_{1}(t)\right] \mathcal{L}%
\left[ f_{2}(t)\right] ,
\end{equation}%
the master equation (\ref{E34}) for the qubit system can be solved as%
\begin{eqnarray}
&&p\tilde{\rho}_{S}^{^{I\prime }}(p)-\rho _{S}^{^{\prime }}(0)  \label{E32}
\\
&=&-A_{-}\sigma _{+}\sigma _{-}\tilde{\rho}_{S}^{^{I\prime }}(p)+\left(
A_{+}+A_{-}\right) \sigma _{-}\tilde{\rho}_{S}^{^{I\prime }}(p)\sigma _{+}
\notag \\
&&-A_{+}\tilde{\rho}_{S}^{^{I\prime }}(p)\sigma _{+}\sigma _{-},  \notag
\end{eqnarray}%
where%
\begin{equation}
A_{\pm }=\sum_{k,i}^{2}\tilde{g}_{k,i}^{2} \;/\left[ p\pm i(\omega
_{k,i}-\eta \,\Delta )\right] .
\end{equation}%
This equation is a Lyapunov matrix equation.

The Kronecker product\ property in matrix theory shows that%
\begin{equation}
\mathrm{Vec}\left( M_{1}\rho M_{2}\right) =M_{1}\otimes M_{2}^{T}\;\mathrm{%
Vec}\left( \rho \right) ,
\end{equation}%
where $\mathrm{Vec}\left( \rho \right) $ represents the vector expanding of
matrix $\rho $ along rows, and the superscript $T$ denotes the transpose of
the matrix. We expand the matrix equation in Eq.~(\ref{E32}) into vectors
along rows:%
\begin{equation}
U(p)\;\mathrm{Vec}\left[ \tilde{\rho}_{S}^{^{I\prime }}(p)\right] =\mathrm{%
Vec}\left[ \rho _{S}^{^{\prime }}(0)\right] ,  \label{E31}
\end{equation}%
where%
\begin{eqnarray}
&&U(p)=pI_{4}+A_{+}I_{2}\otimes \left( \sigma _{+}\sigma _{-}\right) ^{T} \\
&&-(A_{-}+A_{+})\sigma _{-}\otimes \sigma _{+}^{T}  \notag \\
&&+A_{-}(\sigma _{+}\sigma _{-})\otimes I_{2},  \notag
\end{eqnarray}%
where $I_{n}$ is the $n\times n$ identity matrix. Thus, the $2\times 2$
matrix equation (\ref{E32}) is transformed to the $4\times 4$ vector
equation in (\ref{E31}). The solution of Eq.~(\ref{E31}) can be formally
written as%
\begin{equation}
\mathrm{Vec}\left[ \tilde{\rho}_{S}^{^{I\prime }}(p)\right] =U(p)^{-1}\;%
\mathrm{Vec}\left[ \rho _{S}^{^{\prime }}(0)\right] ,
\end{equation}%
with%
\begin{equation}
U(p)^{-1}=\left(
\begin{array}{cccc}
\frac{1}{p+A_{+}+A_{-}} & 0 & 0 & 0 \\
0 & \frac{1}{p+A_{-}} & 0 & 0 \\
0 & 0 & \frac{1}{p+A_{+}} & 0 \\
\frac{1}{p}-\frac{1}{p+A_{+}+A_{-}} & 0 & 0 & \frac{1}{p}%
\end{array}%
\right) .
\end{equation}%
By using the inverse Laplace transform
\begin{equation}
\alpha (t)=\mathcal{L}^{-1}\left[ \tilde{\alpha}(p)\right] =\frac{1}{2\pi i}%
\int\limits_{\sigma -i\infty }^{\sigma +i\infty }dp\;\tilde{\alpha}(p)\exp
\left( pt\right) ,  \label{E33}
\end{equation}%
we obtain,%
\begin{equation}
\mathrm{Vec}[\rho _{S}^{^{I\prime }}(t)]=\mathcal{L}^{-1}U(p)^{-1}\;\mathrm{%
Vec}\left[ \rho _{S}^{^{\prime }}(0)\right] .
\end{equation}%
Then, $\rho _{S}^{^{I\prime }}(t)$ is given by%
\begin{eqnarray}
&&\rho _{S}^{^{I\prime }}(t)  \label{A-12} \\
&=&\left(
\begin{array}{cc}
\mathcal{L}^{-1}\left[ \frac{\rho _{11}(0)}{p+A_{+}+A_{-}}\right] & \mathcal{%
L}^{-1}\left[ \frac{\rho _{12}(0)}{p+A_{+}}\right] \\
\mathcal{L}^{-1}\left[ \frac{\rho _{21}(0)}{p+A_{-}}\right] & \mathcal{L}%
^{-1}\left[ \frac{\rho _{11}(0)}{p}-\frac{\rho _{11}(0)}{p+A_{+}+A_{-}}+%
\frac{\rho _{22}(0)}{p}\right]%
\end{array}%
\right) .  \notag
\end{eqnarray}%
Below we calculate the inverse Laplace transform $\mathcal{L}^{-1}\left(
\frac{1}{p+A_{\pm }}\right) \ $and $\mathcal{L}^{-1}\left( \frac{1}{%
p+A_{+}+A_{-}}\right) $. From Eq.~(\ref{E33}), we have
\begin{equation}
\mathcal{L}^{-1}\left( \frac{1}{p+A_{-}}\right) =\frac{1}{2\pi i}%
\int\limits_{\sigma -i\infty }^{\sigma +i\infty }\frac{\exp (pt)}{%
p+\sum\limits_{k,i}\tilde{g}_{k,i}^{2}/\left[ p-i(\omega _{k,i}-\eta
\,\Delta )\right] }dp.
\end{equation}%
With $p$ replaced by $i\omega +0^{+}$ \cite{I1}, the above expression
becomes
\begin{eqnarray}
&&\mathcal{L}^{-1}\left( \frac{1}{p+A_{-}}\right) \\
&=&\frac{1}{2\pi i}\int\limits_{-\infty }^{+\infty }\frac{\exp (i\omega t)}{%
\omega -\sum\limits_{k,i}\tilde{g}_{k,i}^{2}/\left[ (\omega +\eta \,\Delta
)-\omega _{k,i}-i0^{+}\right] }d\omega .  \notag
\end{eqnarray}%
For the term $\sum_{k}\tilde{g}_{k,i}^{2}/(\omega -\omega _{k,i}-i0^{+}),$
we denote the real and imaginary parts as $R_{i}(\omega )$ and $\Gamma
_{i}(\omega )$, where $i=1$ for the intrinsic bath and $i=2$ for the cavity
bath. Explicitly, we can write%
\begin{eqnarray}
R_{i}(\omega ) &=&\wp \left( \sum_{k}\frac{\tilde{g}_{k,1}^{2}}{\omega
-\omega _{k,1}}\right) \\
&=&\wp \left[ \int_{0}^{\infty }d\omega ^{^{\prime }}\left( \frac{2\eta
_{1}\,\Delta }{\omega ^{^{\prime }}+\eta _{1}\,\Delta }\right) ^{2}\frac{%
J_{1}(\omega ^{^{\prime }})}{(\omega -\omega ^{^{\prime }})}\right] ,  \notag
\end{eqnarray}%
and%
\begin{eqnarray}
\Gamma _{i}(\omega ) &=&\pi \sum_{k}\tilde{g}_{k,i}^{2}\delta (\omega
-\omega _{k}) \\
&=&\pi \left( \frac{2\eta _{i}\,\Delta }{\omega +\eta _{i}\,\Delta }\right)
^{2}J_{i}(\omega ),  \notag
\end{eqnarray}%
where $\wp $ stands for the Cauchy principal value. Let $R(\omega
)=R_{1}(\omega )+R_{2}(\omega )$ and $\Gamma (\omega )=\Gamma _{1}(\omega
)+\Gamma _{2}(\omega ),$ then we have%
\begin{eqnarray}
&&\mathcal{L}^{-1}\left( \frac{1}{p+A_{-}}\right) \\
&=&\frac{1}{2\pi i}\int\limits_{-\infty }^{+\infty }\frac{\exp (i\omega t)}{%
\omega -R(\omega +\eta \,\Delta )-i\Gamma (\omega +\eta \,\Delta )}d\omega .
\notag
\end{eqnarray}%
Similarly, $\mathcal{L}^{-1}\frac{1}{p+A_{+}}\ $and $\mathcal{L}^{-1}\frac{1%
}{p+A_{+}+A_{-}}$ can also be derived as%
\begin{equation}
\mathcal{L}^{-1}\left( \frac{1}{p+A_{+}}\right) =\left[ \mathcal{L}%
^{-1}\left( \frac{1}{p+A_{-}}\right) \right] ^{\ast }  \label{AAA}
\end{equation}%
and \
\begin{widetext}
\begin{eqnarray}
&&\mathcal{L}^{-1}\left( \frac{1}{p+A_{+}+A_{-}}\right)  \\
&=&\frac{1}{2\pi i}\int\limits_{-\infty }^{+\infty }\frac{\exp (i\omega t)}{%
\omega -R(\omega +\Delta \eta )+R(\Delta \eta -\omega )-i\left[ \Gamma
(\omega +\Delta \eta )+\Gamma (\Delta \eta -\omega )\right] }d\omega .
\notag
\end{eqnarray}
\end{widetext}

\section*{Appendix B: Solution of the Schr{\"{o}}dinger equation}

\setcounter{equation}{0} \renewcommand{\theequation}{B.\arabic{equation}}

Below, we will solve the equation of motion of wave function beyond the RWA
in the transformed Hamiltonian $H^{\prime }$ in Eq.~(\ref{e7}). Since the
total excitation number operator of the qubit-cavity system, $%
N=\sum_{k}\left( a_{k}^{\dagger }a_{k}+b_{k}^{\dagger }b_{k}\right) +\left(
1+\sigma _{z}\right) /2$ in the transformed Hamiltonian is a conserved
observable, i.e., $\left[ N,H^{\prime }\right] =0,$ it is reasonable to
restrict our discussion in the single-particle excitation subspace. A
general state in this subspace can be written as
\begin{equation}
\left\vert \Phi (t)\right\rangle =\chi (t)\left\vert s2\right\rangle
\prod\limits_{k}\left\vert 0_{k,1}0_{k,2}\right\rangle +\sum_{k,i}\beta
_{k,i}(t)\left\vert s1\right\rangle \prod\limits_{k}\left\vert 0_{k,%
\overline{i}}1_{k,i}\right\rangle ,
\end{equation}%
where the state $\left\vert 0_{k,\overline{i}}1_{k,i}\}\right\rangle $ means
either cavity bath or qubit spontaneous dissipation bath with one quantum
excitation. Substituting $\left\vert \Phi (t)\right\rangle $ into Schr{\"{o}}%
dinger equation, we have
\begin{equation}
i\frac{d\chi (t)}{dt}=\frac{\eta \,\Delta }{2}\chi
(t)+\sum_{k,i}V_{k,i}\beta _{k,i}(t),  \label{ee1}
\end{equation}%
\begin{equation}
i\frac{d\beta _{k,i}(t)}{dt}=(\omega _{k,i}-\frac{\eta \,\Delta }{2})\beta
_{k,i}(t)+\sum_{k,i}V_{k,i}\chi (t).  \label{ee2}
\end{equation}%
Applying the transformation
\begin{eqnarray}
\chi (t)\! &\!=\!&\!\widetilde{\chi }(t)\exp \left( -i\frac{\eta \,\Delta }{2%
}t\right) ,  \label{e3} \\
\beta _{k,i}(t)\! &\!=\!&\!\widetilde{\beta }_{k,i}(t)\exp \left[ -i(\omega
_{k,i}-\frac{\eta \,\Delta }{2})t\right] ,  \label{e4}
\end{eqnarray}%
Eqs.~(\ref{ee1}) and (\ref{ee2}) is simplified as
\begin{equation}
\frac{d\widetilde{\chi }(t)}{dt}=-i\sum_{k,i}V_{k,i}\widetilde{\beta }%
_{k,i}(t)\exp \left[ -i(\omega _{k,i}-\eta \,\Delta )t\right] ,  \label{e5}
\end{equation}%
\begin{equation}
\frac{d\widetilde{\beta }_{k,i}(t)}{dt}=-iV_{k,i}\widetilde{\chi }(t)\exp %
\left[ i(\omega _{k,i}-\eta \,\Delta )t\right] .  \label{e6}
\end{equation}%
Integrating Eq.~(\ref{e6}) and substituting it into Eq.~(\ref{e5}), we
obtain
\begin{equation}
\frac{d\widetilde{\chi }(t)}{dt}=-\int\limits_{0}^{t}\sum_{k,i}V_{k,i}^{2}%
\exp [-i(\omega _{k,i}-\eta \,\Delta )(t-t^{\prime })]\widetilde{\chi }%
(t^{\prime })dt^{\prime }.  \label{e77}
\end{equation}%
This integro-differential equation~(\ref{e77}) is solved exactly by Laplace
transformation,%
\begin{equation}
\overline{\widetilde{\chi }(p)}=\frac{1}{p+A_{+}}=\frac{\widetilde{\chi }(0)%
}{p+\sum\limits_{k,i}\tilde{g}_{k,i}^{2}/\left[ p-i(\eta \,\Delta -\omega
_{k,i})\right] }.
\end{equation}%
When the initial state is an excited state $\left\vert \psi ^{\prime
}(0)\right\rangle =\left\vert \uparrow \right\rangle \otimes
\prod\limits_{k}\left\vert 0_{k,1},0_{k,2}\right\rangle ,$ i.e., $\widetilde{%
\chi }(0)=1.$ Applying the Inverse Laplace transformation, we get%
\begin{equation}
\widetilde{\chi }(t)=\left( \mathcal{L}^{-1}\frac{1}{p+A_{+}}\right) _{\!t}.
\end{equation}%
From Eq.~(\ref{AAA}), the dynamics evolution of $\rho _{11}^{\prime }$ can
be expressed as%
\begin{eqnarray}
\rho _{11}^{\prime }(t) &=&\chi ^{\ast }(t)\times \chi (t) \\
&=&\left( \mathcal{L}^{-1}\frac{1}{p+A_{+}}\right) _{\!t}\times \left(
\mathcal{L}^{-1}\frac{1}{p+A_{-}}\right) _{\!t}.  \notag
\end{eqnarray}

--------------------

\end{document}